\PassOptionsToPackage{hyphens}{url}
\PassOptionsToPackage{dvipsnames}{xcolor}
\documentclass[conference,compsoc]{IEEEtran}

\usepackage[dvipsnames]{xcolor}
\usepackage[pdftex]{graphicx}
\usepackage[cmex10]{amsmath}
\usepackage{amssymb}
\usepackage{amsthm}
\usepackage{stmaryrd}

\usepackage{enumitem}
\usepackage{cite}
\usepackage{booktabs}
\usepackage{multirow}
\usepackage{caption} 
\usepackage{xspace}
\usepackage{hyperref}
\usepackage[group-separator={,}]{siunitx}
\usepackage[capitalise]{cleveref}

\usepackage[ruled]{algorithm}
\usepackage[noend]{algpseudocode}

\usepackage[scaled=0.88]{helvet}

\usepackage{tikz}
\usetikzlibrary{matrix,shapes,arrows,fit,tikzmark}

\setlist[description]{font=\normalfont\itshape\space}

\newcommand{\Zx}[1]{\mathbb{Z}_{#1}}
\newcommand{\randin}{\gets^{\$}}

\providecommand{\binanswer}{0/1}

\newcommand{\codify}[1]{ \text{\textsf{#1}}\xspace}

\newcommand{\varAsList}[1]{{\langle #1 \rangle}}

\newcommand{\parasec}[1]{\vspace{1mm}\noindent\textbf{#1.}}        
\newcommand{\para}[1]{\vspace{1mm}\noindent\textit{#1.}}                       
\newcommand{\paracodify}[1]{\vspace{1mm}\noindent\codify{#1.}}

\newcommand{\algorithmicSizeModifier}{\small}

\theoremstyle{definition}

\newtheorem*{definition*}{Definition}

\newcommand{\name}{Janus\xspace}

\newcommand{\smcname}{SMC-\name}
\newcommand{\shename}{SHE-\name}
\newcommand{\teename}{TEE-\name}

\newcommand{\rs}{RS\xspace}
\newcommand{\bp}{BP\xspace}
\newcommand{\rsFull}{registration station\xspace}
\newcommand{\bpFull}{biometric provider\xspace}

\def\addvalue#1#2{\expandafter\gdef\csname my@data@#1\endcsname{#2}}%
\def\usevalue#1{\csname my@data@#1\endcsname}

\newcommand{\reqdef}[3]{%
  \addvalue{#2}{RQ.#1}%
  \vspace{1mm}\hypertarget{#2}\noindent\textit{\usevalue{#2}: #3.}
}
\newcommand{\reqlinky}[1]{\hyperlink{#1}{\usevalue{#1}}}

\newcommand{\bioN}{\codify{TS}}
\newcommand{\inputUser}{\codify{recipient}}

\providecommand{\dist}{\codify{Dist}}
\newcommand{\distEuc}{\dist.\codify{Euclidean}}
\newcommand{\distHam}{\dist.\codify{Hamming}}
\newcommand{\distNormHam}{\dist.\codify{NormHamming}}
\newcommand{\createBioSensor}{\codify{Sensor}}
\newcommand{\readBio}{\codify{ReadBio}}
\newcommand{\bioMatch}{\codify{Match}}
\newcommand{\bioAlignedMatch}{\codify{AlignedMatch}}
\newcommand{\bioFusedMatch}{\codify{FusedMatch}}
\newcommand{\bioAlign}{\codify{Align}}

\newcommand{\secretshare}[2]{{#1}^{(#2)}}

\newcommand{\fpprob}{P_{fp}}
\newcommand{\fnprob}{P_{fn}}

\newcommand{\binpowset}[1]{\{0, 1\}^{#1}}
\providecommand{\lxor}{\oplus}

\newcommand{\fheParamGen}{\textsf{HE.ParamGen}}
\newcommand{\fheKeyGen}{\textsf{HE.KeyGen}}
\newcommand{\fheEnc}{\textsf{HE.Enc}}
\newcommand{\fheDec}{\textsf{HE.Dec}}

\newcommand{\encvar}[1]{\llbracket{#1}\rrbracket}
\newcommand{\Zq}{\mathbb{Z}_q}
\newcommand{\param}{\textsf{params}}

\newcommand{\fheaddop}{+}
\newcommand{\fhesubop}{-}
\newcommand{\fhemultop}{\cdot}

\newcommand{\polyDegree}{N_{deg}}
\newcommand{\plainMod}{m_{pt}}
\newcommand{\ctxMod}{m_{ct}}

\newcommand{\bsBlindSign}{\codify{BS.BlindSign}}
\newcommand{\bsSign}{\codify{BS.Sign}}
\newcommand{\bsKeyGen}{\codify{BS.KeyGen}}
\newcommand{\bsVerify}{\codify{BS.Verify}}

\newcommand{\userBio}[1]{X_{#1}}

\tikzset{   
        every picture/.style={remember picture,baseline},
        every node/.style={anchor=base,align=center,outer sep=1.5pt},
        every path/.style={thick},
        }

\newcommand{\scalerFigSingleBio}{0.9}

\begin{document}
    \title{\name: Safe Biometric Deduplication for Humanitarian Aid Distribution}

    \author{
        \IEEEauthorblockN{Kasra EdalatNejad\IEEEauthorrefmark{1}, Wouter Lueks\IEEEauthorrefmark{2}, Justinas Sukaitis\IEEEauthorrefmark{3}, Vincent Graf Narbel\IEEEauthorrefmark{3}\\ Massimo Marelli\IEEEauthorrefmark{3}, Carmela Troncoso\IEEEauthorrefmark{1}}
        \IEEEauthorblockA{
            \IEEEauthorrefmark{1}SPRING Lab, EPFL, Lausanne, Switzerland\\
            \{kasra.edalat, carmela.troncoso\}@epfl.ch
        }
        \IEEEauthorblockA{
            \IEEEauthorrefmark{2}CISPA Helmholtz Center for Information Security, Saarbrücken, Germany\\
            lueks@cispa.de
        }
        \IEEEauthorblockA{
            \IEEEauthorrefmark{3}International Committee of the Red Cross, Geneva, Switzerland\\
            dpo@icrc.org
        }
    }

    \maketitle
    \thispagestyle{plain}\pagestyle{plain}

    \begin{abstract}
        Humanitarian organizations provide aid to people in need. To use their limited budget efficiently, their distribution processes must ensure that legitimate recipients cannot receive more aid than they are entitled to. Thus, it is essential that recipients can register at most once per aid program.

        Taking the International Committee of the Red Cross's aid distribution registration process as a use case, we identify the requirements to detect double registration without creating new risks for aid recipients.
        We then design \name, which combines privacy-enhancing technologies with biometrics to prevent double registration in a safe manner.
        \name does not create plaintext biometric databases and reveals only \emph{one bit} of information at registration time (whether the user registering is present in the database or not).
        We implement and evaluate three instantiations of \name based on secure multiparty computation, somewhat homomorphic encryption, and trusted execution environments.
        We demonstrate that they support the privacy, accuracy, and performance needs of humanitarian organizations.
        We compare \name with existing alternatives and show it is the first system that provides the accuracy our scenario requires while providing strong protection.
    \end{abstract}

\section{Introduction}

Humanitarian organizations have a long history of providing aid to people in crisis. 
To maximize impact, they wish to distribute aid among as many recipients as possible given their limited budget.
Thus, recipients should receive aid only once for each time that aid is distributed.
Wang et al. propose privacy-friendly mechanisms to ensure that recipients enrolled in an aid-distribution system can \emph{receive} aid at most once per round~\cite{Pribad}. Their work relies on the assumption that recipients cannot register more than once.

In this paper, we tackle the problem of \emph{preventing double registration}. To understand the requirements behind double registration prevention in the humanitarian sector, we collaborate with the International Committee of the Red Cross (ICRC). 
We learn that common ways to tackle this issue are to rely on the use of official government-issued identity documents or on the input of local trusted sources of information (e.g., community representatives).

Both methods have shortcomings. Government-issued IDs are not universal. Aid recipients may not have a reliable government identity document, or may not have these documents in their possession at registration time (e.g., after evacuating their homes under pressure in conflict zones). This means that people in need may be refused by the system, opposite to the `humanity' and `impartiality' humanitarian principles stating that aid and protection must be given to everyone that needs it~\cite{icrc-principles}.
Reliance on community representatives may not be available in every scenario, e.g., in dangerous situations. Moreover, it suffers from efficiency issues. Verification by local actors is slow and thus typically can only be done after registration when prevention may not be as useful.
Additionally, both of these methods may increase the risks for the participants in the aid program as they create a single point of failure (e.g., a paper list, a community representative) which, if compromised, reveals the identity of all participants. This membership may be associated with, for example, political or religious beliefs which can be pursued under some governmental regimes~\cite{Afghan-hrw}.

A natural path to address the issues above is digitalization. 
Our conversations with the ICRC reveal that, besides addressing the shortcomings of current proposals, digital solutions need to also minimize the amount of (personal) data collected. 
This is not only to uphold the fundamental right of aid recipients to personal data protection and privacy, but to also help the organization avoid sharing data with third parties (e.g., financial service providers) as such sharing may put recipients in danger especially when those institutions are closely linked with local government or cross-national financial institutions.

A digital means that humanitarian organizations have started to use is biometrics~\cite{RahmanPC18}. 
Biometrics are available in many more situations than identification documents, and the digital nature of biometrics systems helps speed up the registration process.
However, existing biometric-based solutions come at a high risk for recipients.
They require the collection of biometric samples into databases which are typically stored in the clear, becoming a tempting target for entities seeking information about program participants~\cite{Afghan-hrw}. 
Moreover, due to the nature of biometric data, such databases can reveal sensitive information about health or ethnicity~\cite{ElKhiyariW16, RossBC22}, and because biometrics are unique they can be used to link entries across databases. These risks are exacerbated by the long lifespan of biometric data.

Secure and privacy-preserving biometrics solutions~\cite{Tams16, JuelsW99, ImamverdiyevTK13, KumarPR20, YuanY13, BringerCP13, ZhuZXLH18}, which would enable the use of encrypted databases to store the biometric templates, cannot be directly applied to the humanitarian setting. They often do not provide the low error rates required to ensure that legitimate recipients are not erroneously refused from the aid-distribution program. This is because typical error-oriented measures, such as allowing multiple biometric match attempts for authentication, are not a suitable solution for the humanitarian context as in which participants in the matching might want to evade detection.

In this paper, we introduce \name, a biometrics-based deduplication system tailored to the humanitarian sector needs. 
\name only reveals a single \emph{bit} of information determining whether a registration request is valid.
Instead of creating a plaintext biometric database, \name either secret shares biometric data between the registration station and a humanitarian organization-operated biometric provider, encrypt data using a quantum secure HE scheme, or seals the data in a secure enclave.
This approach provides strong privacy protections by ensuring that not even the humanitarian organization has access to plain biometric data; provides long-term protection of sensitive biometric data in the case of data breaches even in the presence of quantum computers; and avoids creating a high-value target for hackers. Finally, \name enables humanitarian organizations to remotely disable query access, even when the registration station is physically compromised during armed conflicts.

Our work makes the following contributions:
\begin{itemize}[nosep, wide]
    \item[\checkmark] We elicit the functionality, safety, and deployment requirements of protecting against double registration in the context of humanitarian aid distribution.
    \item[\checkmark] We propose \name, which addresses these requirements. We build three instantiations of \name based on secure multiparty computation, somewhat homomorphic encryption, and trusted execution environments. We show that \name can support multiple biometric sources including fingerprints, irises, and facial recognition; as well as biometrics alignment and fusion. 
    \item[\checkmark] We implement and evaluate \name's instantiations to demonstrate that they can satisfy the requirements of humanitarian organizations. 
\end{itemize}

\para{Ethical considerations} We do not collect or use biometric data in our evaluation. Therefore, there are no ethical concerns with our experiments.

\section{Deduplication for aid distribution}\label{sec:reqs}
\begin{figure}[tb]
    \centering
    \includegraphics[width=\linewidth, trim={0 0 0 0}, clip]{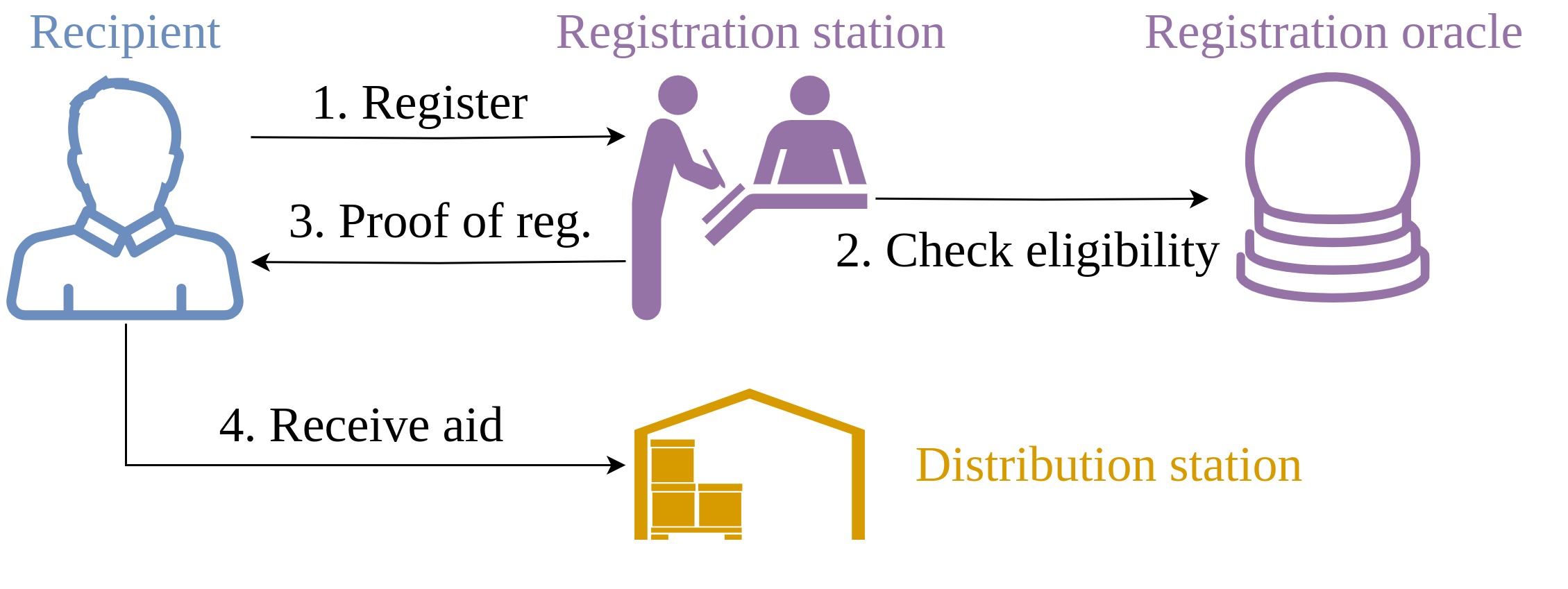}
    \caption{Humanitarian aid distribution workflow~\cite{Pribad}.}
    \label{fig:boyas-parties}
\end{figure}

We take as starting point for our work, the aid distribution scenario identified by Wang et al.~\cite{Pribad}, depicted in \cref{fig:boyas-parties}. This scenario is largely based on the aid distribution process of the International Committee of the Red Cross (ICRC). In the registration phase, potential \emph{aid recipients} (or recipients, for short) visit a \emph{registration station} (RS) and requests aid. If this person is eligible for aid, the registration station enrolls them into the system and allocates them an aid budget. Registration stations are often directly controlled by the humanitarian organization or other trusted local parties. In the case of the ICRC, these stations are protected by the ICRC's special privileges and immunity \cite{BlondCTJFH18}. To ensure the legitimacy of aid recipients, we assume, like Wang et al.~\cite{Pribad}, the existence of a registration oracle that determines eligibility during registration. This role may be fulfilled by verifying recipients' ID documents, checking against local government lists, or asking community representatives. (See \cref{sec:towards-dedup} for why these do not suffice for deduplication.)

At the distribution phase, registered users visit aid distribution centers, show proof of their registration, and receive the aid they are entitled to. As security with respect to the distribution station has been tackled by Wang et al., we leave it out of the scope of this work together with the distribution auditing step included in~\cite{Pribad}.

\subsection{Deduplication Requirements}
\label{sub:requirements}

We collaborate with the International Committee of the Red Cross to understand the requirements associated with preventing double registration, i.e., ensuring that recipients can only register \emph{once} at the registration stations. \looseness=-1
We gather the requirements for deduplication through conversations with the ICRC's Data Protection Office and from ICRC documents on data protection and biometric policy~\cite{data-protection-handbook,icrc-bio-policy}.

\parasec{Functional requirements}
The deduplication mechanism must provide the following functionality:

\reqdef{F1}{iddup}{Identify duplicate aid requests}
The \rsFull needs to determine if a potential recipient is already registered to proceed with a registration request. For this, the deduplication system must provide a \emph{single bit} indicating the potential recipient's presence or absence in the database of registered users.

\reqdef{F2}{dynamic}{Dynamic addition}
If a registration is successful, the newly accepted individual must be added to the registered users database. Newly accepted users must be added individually so that attempts to re-register can be immediately detected. As many recipients may need to be added in a short amount of time, adding individual users to the database should be efficient and not require an operation on all elements in the database.

\reqdef{F3}{lowfailure}{Low failure rate}
Recipients that are eligible should not be erroneously denied registration. Humanitarian organizations should be able to set an arbitrarily low failure rate for the deduplication system on falsely detecting a new recipient as a duplicate.

\para{Non-goals}
The deduplication system only provides information for the \rsFull to make decisions regarding the legitimacy of the registration. 
The deduplication system does not aim to prevent the \rsFull from proceeding with registration in case of a detected duplicate.
For such cases, the humanitarian organization must rely on other processes that ensure that the staff at the \rsFull use the deduplication information correctly.

\parasec{Safety requirements}
Augmenting aid distribution systems with technology may bring new risks to recipients. 
The introduction of deduplication detection must minimize the introduction of such risks.

\reqdef{S1}{single}{Single functionality}
The output of the deduplication system should not reveal more information about recipients than the binary answer necessary to detect duplicates.
Aid recipients often belong to sensitive groups such as refugees or survivors of violent incidents where their data is of interest to various adversaries such as governments.
While some humanitarian organizations (such as the ICRC) enjoy legal protection that prevents subpoenas, creating a valuable database increases the risk for recipients if the data is lost or stolen via hacking attempts~\cite{icrc-hack}.

Therefore, to ensure single functionality, deduplication systems should limit the storage of recipients' personal data, and ensure that any personal data stored cannot be used to recover the identity or any other traits of the recipients. This should also hold for adversaries that have access to other data sources and try to link recipient data to these sources.

\reqdef{S2}{pascomp}{Protection against passive compromise}
Traditional mechanisms for deduplication typically create a single point of failure (e.g., a paper list, a database community representative) which, if compromised, reveals the identity
of all participants and their data. This membership may be associated with, for example, political or religious beliefs which can be pursued under some governmental regimes
To increase safety, a deduplication system should ensure that as long as an adversary compromises at most one actor in the aid distribution system \emph{at the same time}, this adversary cannot learn any information about recipients' data.

\reqdef{S3}{actcomp}{Protection against active compromise}
Humanitarian organizations operate in dangerous and volatile situations, where the \rsFull is at risk of being physically compromised during armed conflicts.
In such scenarios, even the single-bit answer resulting from a membership query for a humanitarian program may put recipients at risk.
For example, when the Taliban took control of Afghanistan, they got access to biometric devices left behind by the US Army.
These devices contained data about Afghan civilians, and the Taliban used them to determine who had a relation to the US Army~\cite{Afghan-hrw}.
Deduplication systems should prevent that third parties can query the system without involving the humanitarian organization.

\parasec{Deployment requirements} Following humanitarian principles, aid programs have to enable humanitarian organizations to serve a large number of users in very diverse environments. We derive the following requirements.

\reqdef{D1}{universal}{Universality}
Humanitarian organizations aim to bring assistance without discrimination. Thus, deduplication systems should only rely on information available in all kinds of contexts, including conflict zones.

\reqdef{D2}{scale}{Medium scalability}
Humanitarian organizations help millions of humans around the world. However, the process of providing aid to recipients is usually localized in geographically-diverse regions where a conflict or disaster happens. Deduplication is only needed within a given aid program in one of these regions.
A typical ICRC program supports \numrange{1000}{10000} recipients. Occasionally, there are larger projects that support up to a \num{100000} recipients. We aim to efficiently operate at the medium scale (around \num{10000}) so that humanitarian organizations can rely on commodity devices and limited network connections, which are typically the only infrastructure available in many of the settings where these organizations operate. To scale to larger projects, humanitarian organizations would require higher computational and communication capabilities, which may limit the scope of application of a solution.

\section{Towards a safe deduplication system}
\label{sec:towards-dedup}
We now discuss methods that humanitarian organizations use, or could use, to prevent duplicate registrations.

\para{Government-issued ID} Government IDs are a simple form of identification. However, they cannot be assumed to be available in humanitarian contexts. Humanitarian organizations often operate in areas there may be no government, or the government refuses to issue IDs to certain groups. Additionally, people fleeing conflict zones often do not prioritize protecting their IDs, and losing them is not uncommon.

\para{Other unique identifiers} An alternative to government IDs is to use government-based (e.g., social security numbers~\cite{Ukraine-ssn}) or commercial-based (e.g. phone numbers or social network aliases) identifiers. These identifiers, however, may not be available everywhere, or it may be easy to obtain several, and given their uniqueness act as (pseudo-)identifiers that can be used to link databases.

\para{Trusted local actors} Humanitarian organizations may rely on trusted local actors, such as local committees, focus rooms, or local community leaders, to identify and verify potential aid recipients. However, this approach is limited by the availability of these trusted actors. Moreover, while local actors are suitable for determining eligibility, humans are not good at remembering who has registered over a long span of time to prevent double registration.

\para{Biographic information} The above measures are sometimes complemented by the use of biographic information to better separate records~\cite{MPDMP}. This can be done using statistical patterns or fuzzy matching, reducing the number of false positives. This method, however, requires the availability of such biographic information, as well as storing it with the corresponding risk.

\para{Biometrics} 
Humanitarian organizations~\cite{RahmanPC18} have started to use human biometrics -- under the assumption that they do not change, and are almost always present -- to implement deduplication while avoiding the weaknesses of the previous methods. For example, the UNHCR introduced the Biometric Identity Management System~\cite{UNHCR-BIMS} and the World Food Program (WFP) uses SCOPE~\cite{WFP-Scope}. Yet, using biometric data brings serious security and privacy concerns. Biometric data are personal, sensitive, and can be used for purposes beyond deduplication. This creates risks for humanitarian organizations in forms of surveillance, hacking attempts, or pressure to share data and risk for aid recipients in forms of losing asylum or extradition~\cite{Rohingya-hrw, Eyes-wide-shut, RahmanPC18, IRIN}. 

Our discussion with the ICRC's Data Protection Office revealed that while government IDs, identifiers, biographic information, and trusted local actors may be suitable for specific aid programs, their lack of availability and accuracy prevents them from fully meeting the functional needs of the ICRC. Therefore, in this work, we focus on biometrics-based solutions combining them with privacy-enhancing technologies to mitigate the risks arising from their use.

\parasec{Biometric-based additional requirements}\label{sec:bio}
Due to their sensitive nature, the use of biometric data requires stricter protection measures to prevent harm. 
For instance, the ICRC's biometrics policy~\cite{icrc-bio-policy} sets strict conditions under which biometric data may be used.
This policy also requires that, in general, biometric materials are stored on users' devices. 
However, this requirement is not compatible with preventing double registrations: we cannot trust malicious recipients to provide biometric data from their previous registrations. Therefore, an effective biometric-based deduplication system must store biometric data on devices operated by the humanitarian organization.

To ensure that the centralized biometric storage does not increase risk (and complies with the biometrics policy) when biometrics are involved, the \emph{single functionality} requirement (\reqlinky{single}) implies that the mechanism must prevent the extraction of the biometric templates (or the recovery of the biometric samples), and any inference on the similarity between two biometric samples. Crucially, the information that the deduplication mechanism requires should not enable the authentication of users.

Finally, given the long lifespan of biometric data and its potential for impersonation and inferring information even decades after collection, we introduce a new requirement: 

\reqdef{S4}{longterm}{Long-term safety}
The deduplication system should be designed to maintain the protection of the information it stores for at least 20-30 years.

\begin{figure}[tb]
    \centering
    \includegraphics[width=\linewidth, clip, trim= 0 0 0 0]{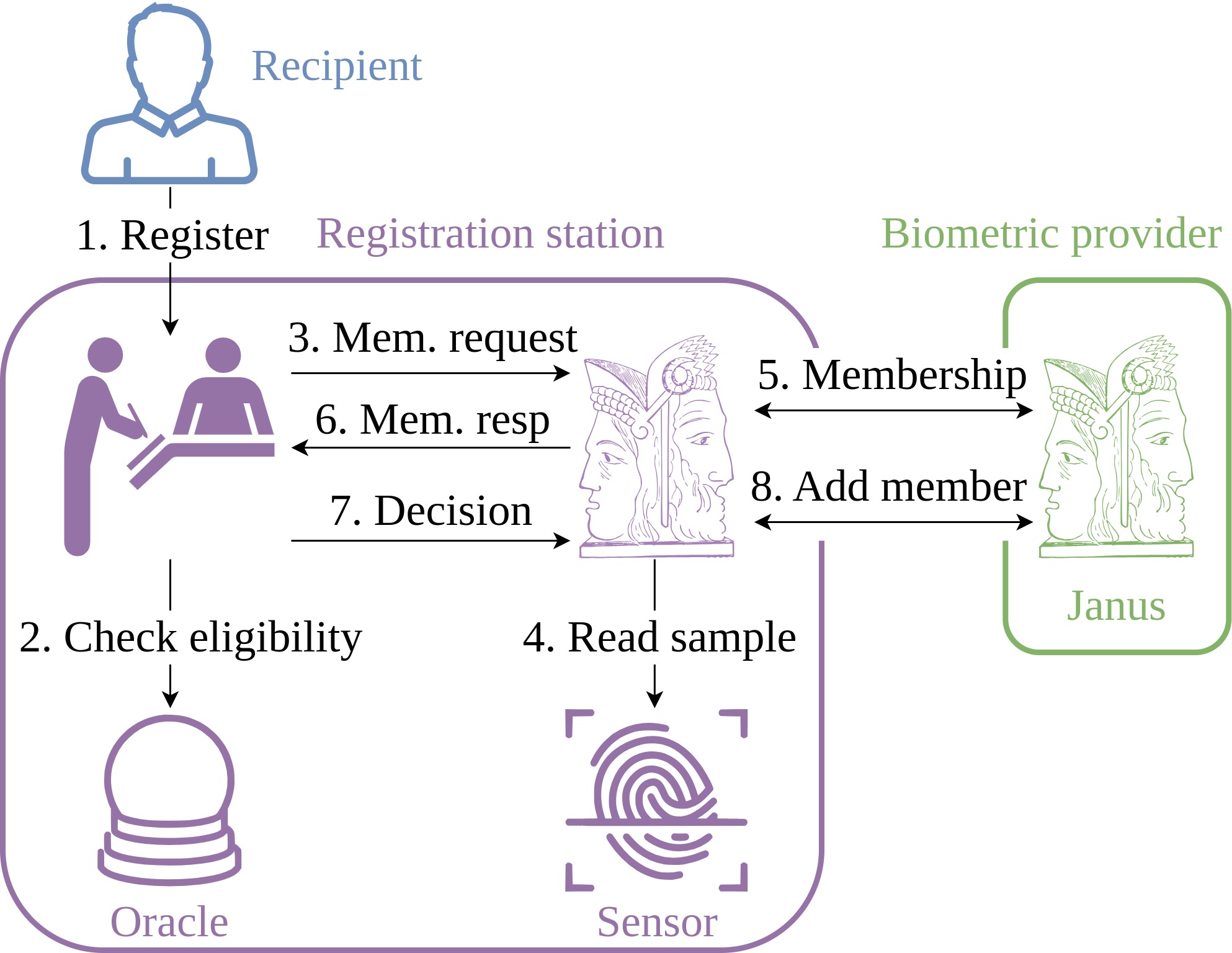}
    \caption{An overview of registration workflow.}
    \label{fig:reg-workflow}
\end{figure}

\section{\name' Architecture}
\label{sec:design}
In this section, we introduce the architecture of \name, a biometric-based deduplication system that fulfills the requirements in Sections \ref{sec:reqs} and \cref{sec:bio}.

\para{Non goals}
\name focuses on preventing double registrations. Our design does \emph{not} aim to: (1) \emph{improve plaintext biometrics} -- we use existing biometric approaches as black boxes; or (2) \emph{secure biometric readings} -- the \rsFull has physical access 
to plaintext biometric samples during registration (via the registration devices). It is trusted to delete them at the end of the registration.
Using biometric sensors with hardware protection would eliminate this need.

\subsection{\name-enabled registration workflow}
In order to fulfill the requirements for protection against passive and active compromise (\reqlinky{pascomp} and \reqlinky{actcomp}), the system must include an extra actor besides the registration station and the recipients (which are malicious in our setting). To this end, we introduce a \emph{biometric provider (\bp)}, an honest-but-curious actor under the control of the humanitarian organization. The \bpFull should not learn any information about aid recipients' biometric data or the success of their registration attempt.
We also consider the \rsFull to be honest-but-curious, and that there is no collusion between the \rsFull and \bpFull.

\newcommand{\inputUserdb}{\codify{db}}
In \cref{sec:deduplication}, we propose three different instantiations of \name' design with different trade-offs. Each of these designs, however, implement the same set of four protocols. A \codify{Setup} procedure initializes the two parties and sets up key material, and initialize a distributed database of registered recipients \inputUserdb. To test for membership of a new recipient in the database \inputUserdb, the two parties jointly run the \codify{Membership} protocol. To add a new recipient to \inputUserdb, the two parties run the \codify{AddMember} protocol. Finally, to provide proactive security, the parties can run the \codify{Ratchet} protocol to recover from earlier compromises of either party.

When including \name to prevent double registration, the registration workflow is as follows (see also \Cref{fig:reg-workflow}):

\begin{enumerate}[noitemsep, topsep=0pt, label=(\roman*)]
    \item A potential aid recipient visits the \rsFull (\rs) to register (step~1). The \rs determines the eligibility and the aid entitlement of the recipient (step~2).
    \item The \rs sends a membership request to the \rs component of \name (step 3). The \rs component then obtains a biometric sample from the recipient (step 4) and runs the \codify{Membership} protocol with the biometric provider component of \name (step 5). \name returns the membership response (step 6).
    \item Upon receiving the membership response from \name, the \rs determines if the registration should proceed or be aborted. If successful (step 7), the \name components in \rs and \bp run the \codify{AddMember} protocol to add the new recipient to the user database (step 8).
    \item Regardless of the outcome, the registration station \emph{deletes} all the plaintext biometric data, and it informs the recipient of the registration's outcome.
\end{enumerate}

From now on, we simply write \rsFull (\rs) to refer to the \rs component of \name and \bpFull (\bp) to refer to the \bp component of \name.

\subsection{Requirements achieved by design}
Our design choices for \name help ensure we meet the requirements in \cref{sec:reqs}.
By design, the interface provided by the $\codify{Membership}$ and $\codify{AddMember}$ protocols satisfy the functional requirements to be able to query for membership (\reqlinky{iddup}) and to add new recipients (\reqlinky{dynamic}). Because \name uses biometrics, it ensures universality (\reqlinky{universal}).

In all instantiations, the separation of trust between the \rs and \bp ensures that \name protects against passive compromises of any single party (\reqlinky{pascomp}). Moreover, because the instantiations support ratcheting, the system can recover from passive compromises of any single party as long as \codify{Ratchet} is run between compromises (\reqlinky{pascomp}).

Our three instantiations of \name differ in the cryptographic building blocks for distributing trust between the \rs and the \bp; which reflects on how they guarantee that each query reveals only a single bit (\reqlinky{single}) and how they achieve scalable designs (\reqlinky{scale}).

\para{Achieving non-collusion}
The non-collusion assumption between the \rs and the \bp is critical to satisfying a number of requirements. In the context of the ICRC, this assumption can easily be realized. The ICRC has a hierarchical structure. The headquarter (HQ) of the ICRC is located in Switzerland, and it has delegations in over 50 countries. The ICRC also has local branches in the 190 countries in which it operates. 

The HQ and majority of delegations are located in physically secure and stable locations and have their own IT infrastructure. Delegations and the HQ are autonomous and suitable for the role of the \bpFull. 
Registration stations need to be closer to potential aid recipients who need to visit the station in person and provide biometric samples. This makes local branches a suitable choice for the role of the \rsFull. The local branches have the IT infrastructure necessary to support gathering and processing biometric samples which enables them to actively participate as a computing party in the deduplication.
This internal structure promotes a significant level of independence between headquarters, delegations, and local branches. With the roles above, the \rs and \bp are typically situated in distinct geographical locations and have separate IT systems; and both benefit from a special privilege of non-disclosure of confidential information~\cite{BlondCTJFH18}, which gives them judicial immunity against subpoenas and requests for data.

\section{Biometrics}
\label{sec:notation}
\label{sub:simplified-bio}
We introduce a simplified and generic abstraction for biometric operations. We first define biometric templates then show how to match two biometric templates to determine if they belong to the same user. We demonstrate that our abstraction supports matching different biometric sources such as fingerprint, iris, and face in \cref{sec:biometric}.

\parasec{Notation}
We write $x \randin X$ to denote that $x$ is drawn uniformly at random from the set $X$. 
Let $q$ be a positive integer, then $\Zq$ denotes the set of integers $[0, \ldots, q)$ and $\Zq^*$ represents the elements of $\Zq$ that are co-prime with $q$. 
We write $\varAsList{a_i}_m$ to represent a list of $m$ elements $[a_0,  \ldots, a_{m-1}]$. 

\parasec{Biometric template} Biometric sensors produce raw images. These images are processed into biometric templates to facilitate matching. We use templates modeled as a \emph{fixed-sized} array of $\bioN$ values taken from an integer domain $\Zx{d}$. We denote the process of reading and processing a biometric sample from a sensor as:
\begin{align*}
S &\gets \createBioSensor(\bioN, d) \\ 
B \in \Zx{d}^{\bioN} &\gets S.\readBio(\inputUser) \text{.} 
\end{align*}
We use $B[i]$ to denote the $i$'th value in a biometric template.
The process of reading biometrics is probabilistic and multiple readings of the same biometric may result in slightly different templates.

\parasec{Biometric matching} Matching takes two biometric templates $X$ and $Y$ and determines if they belong to the same person. We formalize the matching process as: 
\begin{align*}
    \binanswer \gets \bioMatch(X, Y, \dist, t) = (\dist(X, Y) < t) \text{.} 
\end{align*}
where $\dist$ is a distance measure that determines the similarity between two templates, and $t$ is a threshold that determines whether two templates correspond to the same person. 
The choice of the threshold $t$ impacts the error rate of the system and creates a trade-off between false acceptance and rejection (see \cref{sec:biometric}). This parameter is non-sensitive and can be hardcoded into the system.
When the distance and threshold are clear from the context, we drop them from the notation. 

In this work we use three distance measures: \emph{Euclidean distance} ($\distEuc(X, Y) = \sum_i (X[i] - Y[i])^2$), \emph{Hamming distance} ($\distHam(X, Y) = \sum_i (X[i] \lxor Y[i])$), and \emph{normalized Hamming distance} ($\distNormHam((X, M_X),\allowbreak (Y, M_Y))$, where $M_\alpha$ denotes a mask of a template used to exclude certain values during comparison).
We include Euclidean and Hamming distances in the body of the paper and address the normalized Hamming distance in \cref{app:normalized-hamming}.

\parasec{Alignment}
Biometric samples may need alignment before matching. We represent the aligned matching of templates $X$ and $Y$ as producing $a$ aligned templates $\varAsList{X_i}_a \gets \bioAlign(X, a)$ from $X$ then computing pairwise matching between $Y$ and the aligned $X_i$ templates as follows:
\begin{align*}
    M \in \{\binanswer\} \gets &\bioAlignedMatch(X, Y, \dist, t, a) =   \\
    &  \left\{\begin{array}{l}
        \varAsList{X_i}_a \gets \bioAlign(X, a) \\
        d_i \gets \dist(X_i, Y) \\
        M \gets \textstyle \bigvee_{i=1}^{a} (d_i < t) \text{.}\\
    \end{array}\right. 
\end{align*}
\name only supports alignment approaches that do not need to have plaintext access to both $X$ and $Y$ at the same time.
We discuss how to instantiate $\bioAlign$ for different biometric sources in \cref{sec:biometric}.

\section{Instantiating \name}
\label{sec:deduplication}
We design three \name instantiations that use secure multiparty computation, homomorphic encryption, and trusted hardware to offer different trade-offs. These instantiations are inspired by existing biometric systems that we modify to satisfy our specific requirements. 

\newcommand{\rsdb}{\codify{RS.db}}
\newcommand{\bpdb}{\codify{BP.db}}
\newcommand{\dbinsert}{\codify{.Insert}}
\newcommand{\dbgetall}{}
\providecommand{\prf}{\codify{PRF}}
\newcommand{\ssbase}[1]{\codify{SS}_{#1}}
\newcommand{\ssshare}[1]{\ssbase{#1}\codify{.Share}}
\newcommand{\ssreconstruct}[1]{\ssbase{#1}\codify{.Reconstruct}}

\subsection{\smcname}
\label{sub:smc-solution}
Our first instantiation uses secure multiparty computation (SMC) between the \rs and the \bp. We are not the first to propose the use of SMC in the biometrics context. Existing privacy-preserving biometric identification approaches use SMC to compute the distances between a client sample and a database of $N$ samples held by a server~\cite{BringerCP13, BringerCFP0Z14, LuoCPLB12}. These works prevent the server from learning the client's sample and prevent the client from learning the server's database. However, in these approaches the server holds plaintext biometric templates, violating the single functionality requirement (\reqlinky{single}). In our design, we instead store the $N$ samples of registered recipients in secret-shared form between the \rs and \bp. We redesign the SMC-based biometric approach to reconstruct the secret shared templates before computing distances. 

\parasec{Secure two-party computation}
Secure two-party computation is an interactive protocol between Alice with private data $X$ and Bob with private data $Y$ where they want to compute a known function $(A, B) \gets F(X, Y)$ without revealing any information about their private data. The security of S2PC guarantees that the view of each party can be simulated with its input and output. We use Yao's garbled circuit (GC)~\cite{Yao86} for our SMC instantiation.

\parasec{Secret sharing} A secret-sharing scheme shares a secret value $x$ between $n$ party while providing an information-theoretic guarantee that no combination of $t < n$ shares will reveal any information about the secret $x$.
We focus on the two-party scenario, i.e., $n=2$.
A secret-sharing scheme is given by the methods $\ssshare{}$ and $\ssreconstruct{}$.
We define two types of secret sharing, binary and additive:

\vspace{1mm}
\para{$(\secretshare{x}{0}, \secretshare{x}{1}) \gets \ssshare{}(x)$} shares a secret value $x$ between two parties. For binary sharing, $x \in \binpowset{n}$, for additive sharing, $x \in \Zx{\alpha}$. The function $\ssshare{}$ picks the first share $\secretshare{x}{0} \gets \binpowset{n}$ (resp. $\secretshare{x}{0}  \gets \Zx{\alpha}$) uniformly at random in binary (resp. additive) sharing and sets the second share $\secretshare{x}{1}$ such that $\secretshare{x}{0} \lxor \secretshare{x}{1}$ (resp. $\secretshare{x}{0} + \secretshare{x}{1} \mod \alpha$) equals $x$. We write $\ssbase{b}$ to denote binary secret sharing and $\ssbase{\alpha}$ to denote arithmetic sharing modulus $\alpha$.

\vspace{1mm}
\para{$x \gets \ssreconstruct{}(\secretshare{x}{0}, \secretshare{x}{1})$} takes two shares and outputs the original secret $x = \secretshare{x}{0} \lxor \secretshare{x}{1}$ in the binary and $x = (\secretshare{x}{1} + \secretshare{x}{0}) \bmod \alpha$ in the additive setting.

\parasec{\smcname} We now instantiate the \smcname functions.

\paracodify{Setup} We initialize the system as follows:
\begin{enumerate}[noitemsep, topsep=0pt]
    \item The \rs initializes a new biometric sensor $S \gets \createBioSensor(\bioN, d)$.
    \item The \rs creates an empty database $\rsdb$ for storing secret shared biometric templates.
    \item The \bp creates an empty database $\bpdb$ for storing secret shared biometric templates.
\end{enumerate}

\paracodify{Membership}
The \rs reads a biometric sample $B$ from the new recipient, then interacts with the \bp to compute whether it matches any of the $N$ registered recipients in the database using a garbled circuit:
\begin{enumerate}[noitemsep, topsep=0pt]
    \item The \rs reads a new biometric sample from the recipient $B \gets S.\readBio(\inputUser)$.
    \item The \rs computes $a$ alignments of $B$ using $\varAsList{B_j}_a \gets \bioAlign(B, a)$.
    \item The \rs and \bp use a garbled circuit to compute the $\codify{SMC-Matching}$ function from \cref{alg:smc-gc} as follows:
    \begin{enumerate}
        \item The \rs retrieves the registered users' template shares from its database $\varAsList{\secretshare{X_i}{0}}_N \gets \rsdb\dbgetall$ and uses $(\varAsList{\secretshare{X_i}{0}}_N, \varAsList{B_j}_a)$ as its private input.
        \item  The \bp retrieves the secret shares of registered users' templates $\varAsList{\secretshare{X_i}{1}}_N \gets \bpdb\dbgetall$ as its private input.
        \item The \rs and \bp jointly compute $\codify{SMC-Matching}$ and \rs learns the binary value $\codify{member}$.
    \end{enumerate}
    \item The \rs outputs the binary value  $\codify{member}$.
\end{enumerate}
When computing the $\codify{SMC-Matching}$ function, we first reconstruct biometric templates for registered users in line 3. Next, we compute the Hamming distance between the $i$'th user and the $j$'th alignment of $B$ in line 5. We check the distance against the public threshold $t$ to decide if a pair of templates are a match in line 6. The user $i$ is a match for the new recipient if the template $X_i$ matches any of the $a$ alignment $B_j$ as seen in line 7. Finally, recipients are duplicates if they match any of the existing users (line 8). Changing the Hamming distance to Euclidean distance only requires replacing line 5 with $d_{i, j} \gets \sum_{k=0}^{\bioN-1} (B_j[k] - X_i[k])^2$.

\paracodify{AddMember} To add a new recipient, the \rsFull loads the biometric from the pending registration request and secret shares the template $B$ with the \bpFull (without including $a$ alignments) as follows:
\begin{enumerate}[noitemsep, topsep=0pt]
    \item The \rs loads the plaintext biometric template $B$ from the last membership request and secret shares $B$ as $(\secretshare{B}{0}, \secretshare{B}{1}) \gets \ssshare{b}(B)$.
    \item The \rs runs $\rsdb\dbinsert(\secretshare{B}{0})$ to insert $\secretshare{B}{0}$ into its database and sends $\secretshare{B}{1}$ to \bp. The \bp inserts $\secretshare{B}{1}$ into its the user database by running $\bpdb\dbinsert(\secretshare{B}{1})$.
    \item The \rs deletes plaintext biometrics templates $B$, alignments $ \varAsList{B_j}_a$, and \bp's share $\secretshare{B}{1}$.
\end{enumerate}

\begin{algorithm}[tb]
    \caption{The matching functioned computed using a garbled circuit inside \smcname (for hamming distances).}
    \label{alg:smc-gc}
    \begin{algorithmic}[1]
        \algorithmicSizeModifier 
        \Function{SMC-Matching}{$(\varAsList{\secretshare{X_i}{0}}_N, \varAsList{B_j}_a), \varAsList{\secretshare{X_i}{1}}_N$}
            \For {$i \gets \Zx{N}$ }
                \State  $X_i \gets \secretshare{\userBio{i}}{0} \lxor \secretshare{\userBio{i}}{1}$
                \For{ $j \gets \Zx{a}$}
                    \State $d_{i, j} \gets \sum_{k=0}^{\bioN-1} (B_j[k] \lxor X_i[k])$ 
                    \State $m_{i, j} \gets d_{i, j}  < t $ 
                \EndFor
                \State $\codify{match}_i  \gets \bigvee_{j=0}^{a-1} m_{i, j}$
            \EndFor
            \State $\codify{member} \gets \bigvee_{i=0}^{n-1} \codify{match}_i$
            \State \Return $\codify{member}$ 
        \EndFunction

    \end{algorithmic}
\end{algorithm}

\paracodify{Ratchet}
Ratcheting refreshes the secret shares of the \rsFull and \bpFull (see \reqlinky{pascomp}).
To minimize communication costs, instead of sending new shares for all values, we send a single randomness seed and use a keyed PRF function to refresh shares locally as follows:
\begin{enumerate}[noitemsep, topsep=0pt]
    \item The \rs chooses a randomness seed $r$ and sends it to the \bpFull.
    \item The \bp receives the seed $r$ and randomizes template shares $\varAsList{\secretshare{X_i}{1}}_N  \gets \bpdb\dbgetall$ as $\secretshare{X'}{1}_i  \gets \secretshare{X'}{1}_i  \lxor \prf_r(i)$.
    The \bp deletes its database $\bpdb$ and the seed $r$ then create a new user database with new shares using $\bpdb\dbinsert(\varAsList{\secretshare{X'_i}{1}}_N )$.
    \item The \rs refreshes its shares using the seed $r$ following the same process as \bp. Afterward, \rs deletes the seed $r$ and all old secret shares.
\end{enumerate}

\parasec{Safety} The SMC-based design satisfies the safety requirements. First, we argue that \smcname achieves single functionality (\reqlinky{single}). By inspection, the only protocol that might reveal data related to recipients is \codify{Membership} (at \codify{Setup} there is no data to leak, and \codify{AddMember} and \codify{Ratchet} have no output). However, in \codify{Membership} the private inputs of either party are fed into a two-party SMC protocol that guarantees that as long as the parties are honest-but-curious (which they are by assumption), only the \rs learns the single output bit, and nothing more. Thus \reqlinky{single} is satisfied.

Second, at rest, recipient-related data is secret-shared between the \rs and the \bp, ensuring information-theoretic protection against single compromises. As a result, a compromise of either party does not reveal any sensitive data (satisfying \reqlinky{pascomp}) and long-term safety is guaranteed (\reqlinky{longterm}). Any leakage of information resulting from queries can only be the result of a call to $\codify{Membership}$ by an attacker compromising the \rs. However, the SMC protocol can only be run with the cooperation of the \bp (satisfying \reqlinky{actcomp}).

\subsection{\shename}
\label{sub:fhe-solution}
Next, we instantiate \name using homomorphic encryption.
Biometric matching and identification can be computed using either additively homomorphic encryption~\cite{HuangMEK11, BarniBCRLFFLPSP10} or fully homomorphic encryption~\cite{KumarPR20, YasudaSKYK15}. Computing the Euclidean distance is straightforward with homomorphic encryption. However, many HE schemes are not suitable for performing the comparison needed for applying the decision threshold.
While many of the prior works~\cite{Boddeti18, EngelsmaJB22} reveal this distance and perform the thresholding in the clear, some papers like Huang et al.~\cite{HuangMEK11} extend the HE distance computation with an SMC comparison to keep the whole computation in an encrypted domain. 
We use the BFV~\cite{FanV12} somewhat homomorphic encryption to compute Euclidean distance and use a garbled circuit to perform the comparison. 
We compare our performance against a scheme that reveals the distance~\cite{EngelsmaJB22} and Huang et al.~\cite{HuangMEK11} that protects the distance in \cref{sub:comparison}.

\parasec{Somewhat homomorphic encryption}
Somewhat homomorphic encryption schemes allow arithmetic operations like additions and limited multiplications without decryption. This paper relies on the BFV scheme over the prime ring $\Zq$.
There are 4 major methods in SHE schemes:
\begin{itemize}[noitemsep, topsep=0pt]
\item $\param \leftarrow \fheParamGen(q)$. Takes a plaintext domain $\Zq$ and generates an HE parameter set.
\item $pk, sk \leftarrow \fheKeyGen(\param)$. Generates a new key pair $(pk, sk)$ based on the parameter set $\param$. Evaluation keys are assumed to be part of the public key.
\item $\encvar{x} \leftarrow \fheEnc(pk, x)$. Takes the public key $pk$ and a message $x \in \Zq$ and returns the ciphertext $\encvar{x}$.
\item $x \leftarrow \fheDec(sk, \encvar{x})$. Takes the secret key $sk$ and a ciphertext $\encvar{x}$ and returns the decrypted message $x$.
\end{itemize}
The correctness property of the encryption ensures that $\fheDec(sk, \fheEnc(pk, x)) \equiv x \pmod q$.

\para{Homomorphic operations}
SHE schemes support homomorphic addition (denoted by~$\fheaddop$), subtraction (denoted by~$\fhesubop$), and a limited number of multiplication (denoted by~$\fhemultop$) of ciphertexts: $\fheDec(sk, \encvar{a} \fhemultop \encvar{x} \fheaddop \encvar{b}) = ax + b \bmod q$.
It is also possible to perform addition and multiplication with scalar values in addition to operating on two ciphertexts.

\para{SIMD batching}
We can combine number theoretic transformation (NTT) with the BFV scheme to allow batching $N$ scaler values into each ciphertext~\cite{SmartV14}. This transforms each arithmetic operation to a SIMD alternative that performs pairwise operations between two $N$-ary vectors.

\begin{algorithm}[tb]
    \caption{The matching functioned computed using a garbled circuit inside \shename.}
    \label{alg:she-gc}
    \begin{algorithmic}[1]
        \algorithmicSizeModifier 
        \Function{SHE-Matching}{$\varAsList{\secretshare{d_{i, j}}{0}}_{aN}, \varAsList{\secretshare{d_{i, j}}{1}}_{aN}$}
            \For {$i \gets \Zx{N}$ }
                \For{ $j \gets \Zx{a}$}
                    \State $d_{i, j} \gets \secretshare{d_{i, j}}{0} + \secretshare{d_{i, j}}{1}$
                    \If {$d_{i, j} \geq q$}
                        \State $d_{i, j} \gets d_{i, j} - q $
                    \EndIf
                    \State $m_{i, j} \gets d_{i, j}  < t $ 
                \EndFor
                \State $\codify{match}_i  \gets \bigvee_{j=0}^{a-1} m_{i, j}$
            \EndFor
            \State $\codify{member} \gets \bigvee_{i=0}^{n-1} \codify{match}_i$
            \State \Return $\codify{member}$ 
        \EndFunction

    \end{algorithmic}
\end{algorithm}

\parasec{\shename} We now instantiate the \shename functions.

\paracodify{Setup} We initialize the system as follows:
\begin{enumerate}[noitemsep, topsep=0pt]
    \item The \rs initializes a new biometric sensor following $S \gets \createBioSensor(\bioN, d)$.
    \item The \bp creates a new SHE key pair $pk, sk \leftarrow \fheKeyGen(\param)$ and sends the public key to the \rsFull. The \rs saves $pk$.
    \item The \rs creates an empty database $\rsdb$ for encrypted biometric templates.
\end{enumerate}

\paracodify{Membership} To determine if a recipient has registered before, the \rsFull reads a biometric sample $B$ and performs the membership matching in three phases: computing Euclidean distance using SHE, secret sharing the encrypted distance using an additive sharing modulus $q$, and using a GC to perform comparison over secret shared values.
\begin{enumerate}[noitemsep, topsep=0pt]
    \item The \rs reads a new biometric sample from the recipient $B \gets S.\readBio(\inputUser)$ and computes $a$ alignments of $B$ using $\varAsList{B_j}_a \gets \bioAlign(B, a)$.
    \item  The \rs retrieves the encrypted templates of registered recipients $\varAsList{\encvar{X_i}}_N \gets \rsdb\dbgetall$ from its database.
    \item The \rs computes the Euclidean distance of the alignment $B_j$ and user template $\encvar{X_i}$ as $\encvar{d_{i, j}} =  \sum_{k=1}^{\bioN}(\encvar{X_i[k]} - B_j[k])^2$.
    \item The \rs secret shares the encrypted Euclidean distances using $\secretshare{d_{i, j}}{0} \randin \Zq, \encvar{\secretshare{d_{i, j}}{1}} \gets  \encvar{d_{i, j}} - \secretshare{d_{i, j}}{0}$ and sends $\varAsList{\encvar{\secretshare{d_{i, j}}{1}}}_{aN} $ to the \bpFull.
    \item The \bp receives and decrypts $\varAsList{\encvar{\secretshare{d_{i, j}}{1}}}_{aN}$. 
    \item The \rs and \bp interact together to compute the $\codify{SHE-Matching}$ function in \cref{alg:she-gc} using a garbled circuit as follows:
    \begin{enumerate}
        \item The \rs uses $(\varAsList{\secretshare{d_{i, j}}{0}}_{aN})$ as its private input.
        \item The \bp uses $(\varAsList{\secretshare{d_{i, j}}{1}}_{aN})$ as its private input.
        \item The \rs and \bp jointly compute the garbled circuit and \rs learns the value $\codify{member}$ as output.
    \end{enumerate}
    \item The \rs outputs the binary value  $\codify{member}$.
\end{enumerate}

Computing the Euclidean distance $\encvar{d_{i, j}}$ between the $j$'th alignment $B_j$ and $i$'th user template $\encvar{X_i}$ is straightforward and follows the definition of Euclidean distance. The Hamming distance is equivalent to Euclidean distance in binary arrays; we discuss an optimized version of Hamming distance for our HE approach in \cref{app:normalized-hamming}.
We use additive secret sharing modulo $q$ to secret share the distance. When performing the secret sharing over the ciphertexts, the BFV scheme automatically performs the modular reduction. However, we need to manually perform the modular reduction needed for reconstruction in the GC. Instead of running a costly modulus operation in GC, we take advantage of the fact that both shares $\secretshare{d_{i, j}}{0}, \secretshare{d_{i, j}}{1}$ are less than $q$ and consequently their addition is less than $2q$. This allows us to simplify computing modulus in line 6 to subtract $q$ if $d_{i, j}$ is greater than $q$. After computing $d_{i, j}$ in our garbled circuit protocol, comparing against the decision threshold is straightforward and similar to our SMC approach.

\paracodify{AddMember} To add a new recipient, the \rsFull encrypts then stores the the resulting ciphertext template in $\rsdb$ as follows:
\begin{enumerate}[noitemsep, topsep=0pt]
    \item The \rs loads the plaintext templates $B$ from the last membership request,
    encrypts $\encvar{B} \gets \fheEnc(sk, B)$, and inserts it to the user database $\rsdb\dbinsert(\encvar{B})$.
    \item The \rs deletes all plaintext and alignment templates $B, \varAsList{B_j}_a$.
\end{enumerate}

\paracodify{Ratchet} The \bpFull generates a new key pair then interacts with the \rs to transition the \rsFull's old ciphertext to the new key as follows:
\begin{enumerate}[noitemsep, topsep=0pt]
    \item The \bp generates a new key pair $pk', sk' \gets \fheKeyGen(\param)$ and sends the public key $pk'$ to the \rsFull.
    \item The \rs receives and saves the public key $pk'$.
    \item In a similar manner to membership, the \rs secret shares the templates  $\secretshare{X_i}{0} \randin \Zq, \encvar{\secretshare{X_i}{1}} \gets  \encvar{X_i} - \secretshare{X_i}{0}$ and sends $\varAsList{\encvar{\secretshare{X_i}{1}}}_N$ to the \bpFull.
    \item The \bp receives $\varAsList{\encvar{\secretshare{X}{1}}}_N$, decrypts then re-encrypt template shares with the new key $pk'$ as $\encvar{\secretshare{X'_i}{1}} \gets \fheEnc(pk', \fheDec(sk, \encvar{\secretshare{X_i}{1}}))$.
    \item The \bp sends $\varAsList{\encvar{\secretshare{X'_i}{1}}}$ to the \rsFull.
    \item The \rs deletes the old ciphertext database and then reconstructs and inserts new ciphertext to the database as $\rsdb\dbinsert(\varAsList{\encvar{\secretshare{X'_i}{1}} + \secretshare{X_i}{0}}_N)$.
    \item The \bp deletes the old key pair $(pk, sk)$.
\end{enumerate}

\parasec{Safety}
Our SHE-based version satisfies the safety requirements.
Similar to \smcname, the only protocol that reveals an output about recipients is \codify{Membership}.
The \codify{Membership} protocol has three stages. First, the \rsFull computes the distance in the encrypted domain, thus cannot gain any information. Second, \rs secret shares the distances and sends shares $\secretshare{d_{i, j}}{1}$ to the \bp, thus the security of $\ssbase{}$ prevents any leakage. Third, both parties feed their secret shares into a two-party SMC protocol that guarantees that as long as the parties are honest-but-curious (which they are by assumption), only the \rs learns the single output bit. Therefore, \reqlinky{single} is satisfied.

At rest, the \bpFull does not have any information about recipients, and the \rsFull stores a database encrypted with a quantum secure encryption scheme without knowing the decryption key. As long as only the data of a single party gets compromised during each ratchet epoch, \shename satisfies both \reqlinky{pascomp} and \reqlinky{longterm}. Similar to \smcname, the leakage from queries is limited by the need for cooperation from the \bp to run calls to $\codify{Membership}$ (satisfying \reqlinky{actcomp}).

\subsection{\teename}
\label{sub:tee-solution}
An alternative for using advanced cryptographic tools such as SMC and SHE is relying on trusted hardware. In this section, we instantiate \name based on a trusted execution environment (TEE). While TEEs allows us to ensure the security of biometric templates with only one party, we still keep our two-party setting to protect against active compromises (\reqlinky{actcomp}). In \teename, we use the \bpFull as an authentication server that blindly signs membership requests. If an adversary takes control of the \rsFull, the humanitarian organization can disable the \bpFull to prevent adversaries from performing queries and learning information from \rs.

Recent attacks on trusted hardware~\cite{KimDKFLLWLM14, Lipp0G0HFHMKGYH18, BulckMWGKPSWYS18} demonstrate that extracting secrets from a TEE is not impossible. 
While we acknowledge the vulnerability of TEE solutions to hardware attacks as a limitation of \teename, our design safeguards against side channels like memory access and timing patterns.
We deploy TEEs on humanitarian organization-controlled devices and not on a cloud service, therefore, there is no need for TEE attestation.

\parasec{Trusted execution environment} TEEs are hardware security modules that ensure the confidentiality and integrity of the data and code during execution. TEE programs have two components: an enclave that runs the sensitive portion of the code and benefits from the hardware security guarantee, and a host component that runs outside of the secure module and is in charge of managing the interactions of the enclave with non-secure components (e.g., the network).

\parasec{Blind signature}
A blind signature scheme allows a client to interact with a server to compute a signature on private message $m$ with the server's secret key without revealing $m$ to the server.
There are 4 major methods in blind signatures:

\begin{itemize}[noitemsep, topsep=0pt]
\item $vk, sk \gets \bsKeyGen()$. Generates a verification key $vk$ and a signing key $sk$.
\item $\sigma \gets \bsSign(sk, m)$. The server signs a known message $m$ with its signing key $sk$.
\item $\sigma \gets \bsBlindSign(sk, m)$. A client with a secret message $m$ interacts with a server holding the signing key $sk$ to compute a signature $\sigma$ on $m$. Blindness guarantees that the server does not learn any information about the message.
\item $\binpowset{} \gets \bsVerify(vk, m, \sigma)$. Takes a message $m$, a signature $\sigma$, and a verification key $vk$ and determines if $\sigma$ is a valid signature for $m$.
\end{itemize}

\parasec{\teename} We now instantiate the \teename functions.

\paracodify{Setup} We initialize the system as follows:
\begin{enumerate}[noitemsep, topsep=0pt]
    \item The \bp creates a new blind signature key $vk, sk \gets \bsKeyGen()$ and sends the verification key $vk$ to the \rsFull.
    \item The \rs creates a secure enclave and stores $vk$ as the authority key inside. The enclave securely creates an empty database $\rsdb$ for biometric templates.
    \item The \rs initializes a new biometric sensor through the host following $S \gets \createBioSensor(\bioN, d)$. Remember that running the sensor inside an enclave is a non-goal for this work (see \cref{sub:requirements}).
\end{enumerate}

\paracodify{Membership} To check if a new user is already registered in the database, the \rsFull reads and processes a biometric sample $B$ on the host, requests a blind signature from the \bpFull on this template, and passes both the template and signature to the enclave which can directly perform the plaintext biometric matching after verifying the signature.
\begin{enumerate}[noitemsep, topsep=0pt]
    \item The \rs reads a new biometric sample from the recipient $B \gets S.\readBio(\inputUser)$.
    \item The \rs host interacts with the \bp to compute a blind signature on the template $\sigma \gets \bsBlindSign(sk, B)$, and passes passes $(B, \sigma)$ to the enclave.
    \item The \rs enclave verifies the signature with the stored authority key $vk$ by running $\bsVerify(vk, B, \sigma)$. If the verification fails, the enclave aborts.
    \item The \rs enclave runs a plaintext biometric membership check by iterating over \emph{all} registered templates and comparing them with the plain template $B$ to compute the binary result $\codify{member}$. Comparing against all templates prevent memory access pattern and timing attacks by ensuring that the operations performed are independent of the data.
    \item The \rs outputs the membership result $\codify{member}$.
\end{enumerate}

\paracodify{AddMember} To add a new recipient, the \rs inserts the biometric sample from the last membership query to the sealed registered user database inside the enclave as follows:
\begin{enumerate}[noitemsep, topsep=0pt]
    \item The \rs enclave loads the plaintext biometric templates $B$ from the last membership request and inserts the template to its user database $\bpdb\dbinsert(\encvar{B})$.
    \item The \rs host deletes the plaintext templates $B$.
\end{enumerate}

The \rsFull is adding the biometric template from the last membership request to the database without reading a new sample. As \rs has already received and verified a blind signature on this template, there is no need for repeating the blind signing.

\paracodify{Ratchet}
We assume that the biometric data stored in the enclave are secure and the adversary cannot extract secrets from a TEE. Therefore, our ratchet protocol only supports rotating the key of the \bpFull after verifying the signature of the ratchet request as follows:
\begin{enumerate}[noitemsep, topsep=0pt]
    \item The \bpFull generates a new key pair $vk', sk' \gets \bsKeyGen()$, signs the verification key with the old secret key $\sigma \gets \bsSign(sk, vk')$, and sends the verification key $vk'$ and the signature $\sigma$  to the \rsFull.
    The \rs host receives the new key and signature $(vk', \sigma)$ and passes them to the enclave.
    \item The \rs enclave loads the old verification key $vk$ from the sealed memory and runs $\bsVerify(vk, vk', \sigma)$.
    If the verification passes, the \rs enclave replaces the old verification key $vk$ with the new one $vk'$.
\end{enumerate}

\parasec{Safety}
TEE-\name satisfies the safety requirements. First, we argue that the design achieves single functionality (\reqlinky{single}). All recipient data is stored inside the TEE enclave. The only time that the enclave reveals any information about these data is when running \codify{Membership}. By design, the enclave code will compute and output exactly the allowed 1-bit membership response, thus satisfying \reqlinky{single}.

Second, the sealing properties of the TEE guarantee security against passive compromises of either party, thus satisfying \reqlinky{pascomp}. Third, the TEE only replies with a 1-bit answer when it receives a (blindly) signed request from the \bp. Thus, TEE-Janus also protects against active compromises, satisfying \reqlinky{actcomp}.
Finally, provided that the security guarantees of TEEs hold, TEE-Janus provides long-term safety (\reqlinky{longterm}). However, given the recent attacks on TEEs~\cite{KimDKFLLWLM14, Lipp0G0HFHMKGYH18, BulckMWGKPSWYS18} we do not want to claim that TEE-Janus provides long-term safety in practice.

\section{Biometrics in practice}
\label{sec:biometric}
Biometric pipelines start with a raw image such as an eye image or a fingerprint scan. 
Directly comparing images is complex and costly. Thus, typically images are transformed into lower-dimension representations called `templates'. 
This transformation can rely on classic algorithms~\cite{JainPHP99, tisseMTR02,yang11} or on neural networks~\cite{DengGYXKZ22, SchroffKP15, BarzutMASMG21}.
As we assume the \rsFull processes biometrics in plaintext, \name can use prior biometric works as a black-box inside the $\readBio$ function,
as long as the resulting template complies with the abstraction in \cref{sub:simplified-bio}. 

\parasec{Template creation} We show how to generate templates and handle alignment for major biometric sources. We note that when choosing biometrics, humanitarian organizations must consider factors such as culture and religion which may vary between regions where aid needs to be distributed.

\para{Fingerprint} The traditional approach for fingerprint matching is minutiae matching. Unfortunately, it does not satisfy our abstraction as the number of extracted minutiae varies between readings and minutiae do not have a fixed ordering. Thus, we use a different matching technique relying on \emph{finger codes}~\cite{JainPHP99, JainPHP00, ShaZT03, EngelsmaCJ21}, that transform fingerprints into fixed-size arrays of integers where similarity is computed based on Euclidean distance.
Different fingerprint scans can have different angles. Some finger code approaches~\cite{JainPHP99, ShaZT03} create codes that are not rotation-invariant. Instead, they use an alignment process that rotates the raw image $a$ times (commonly by $11.5^{\circ}$ degree), extract $a$ finger codes, and check if any of these $a$ templates match the target biometric to improve accuracy.

\para{Iris} Irises are a highly accurate source of biometric data. A common matching approach is \emph{iris code}~\cite{Daugman04, hollingsworthBF08, tisseMTR02}, where an eye image is converted to a fixed-size binary array.
Not all iris bits are equal. Parts of the eye may be obstructed (e.g., by glares or eyelashes) and lead to unreliable bits~\cite{hollingsworthBF08}. To address this issue, iris codes are accompanied by mask vectors to exclude unreliable bits from the comparison, and similarity is computed using normalized Hamming distance.
Iris codes have an alignment procedure that applies $a$ circular shifts (by steps of 1 or 2) on the bit vector and repeats the similarity comparison. 

\para{Face} Prior works such as ArcFace~\cite{DengGYXKZ22} allow us to represent faces with a fixed-size array of real values where the similarity of two faces is computed using Euclidean distance. We can discretize these real values into integers fitting into an arbitrary domain $\Zx{d}$. This discretization can have a minor impact on accuracy that depends on the choice of $d$. The templates produced in ArcFace do not require alignment.

\subsection{Membership with a single sample}
We study membership performance with respect to two types of errors, which we name following the same convention as in biometric identification and matching:  \emph{false accept rate (FAR)}, the probability of incorrectly identifying a new person as an existing user in a biometric membership; and \emph{false reject rate (FRR)}, the probability of failing to detect an existing user in a biometric membership test.

In the humanitarian setting, the cost of false acceptance is considerably higher than false rejection. False acceptances prevent legitimate recipients from receiving aid. False rejection enables recipients to register more than once. While allowing cheating is undesirable, humanitarian principles imply that we must favor lower FAR over FRR (see \reqlinky{lowfailure}).

The membership operation can be broken down into individual matching operations between the queried biometric sample $B$ and the $N$ templates of recipients already registered in the system. The accuracy of each matching operation is limited by the error rate of the plaintext matching operation $\bioMatch$. Assuming that the underlying plaintext matching has a false positive rate of $\fpprob$ and a false negative rate of $\fnprob$, and these probabilities are independent for different match operations, we can compute FAR as $1 - (1 - \fpprob)^N$ and FRR as $\fnprob \left(1 - \fpprob\right)^{N-1}$.

\begin{figure}[t]
    \centering
    \includegraphics[width=\scalerFigSingleBio\linewidth, clip]{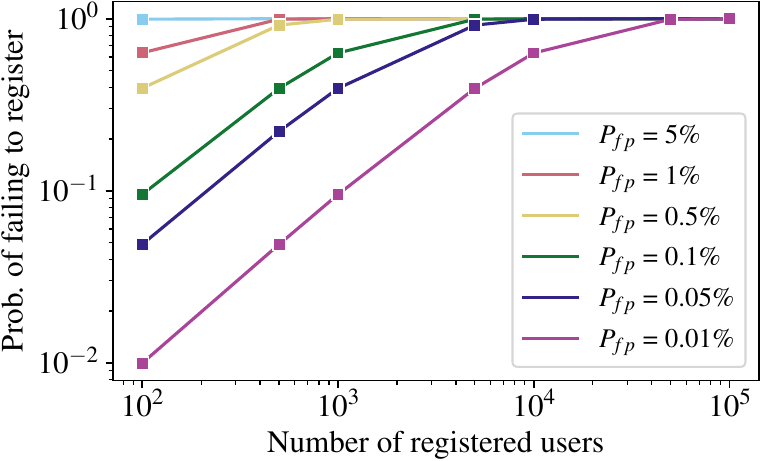}
    \caption{Impact of biometric matching $\fpprob$ on registration.}
    \label{fig:far-impact}
\end{figure}

The error probabilities of plaintext matching algorithms for a single biometric sample are typically in the range of $0.01-5\%$.
These error rates often satisfy the accuracy requirements of authentication systems and choosing the decision threshold value $t$ allows configuring a trade-off between $\fpprob$ and $\fnprob$. These rates however are not sufficient in the context of humanitarian aid distribution. \Cref{fig:far-impact} shows the probability of \emph{falsely} rejecting a legitimate new recipient depending on the number of registered users and biometric failure probability $\fpprob$. Looking at this graph, deduplicating a new recipient with a single biometric sample against a database of \num{10000} fails with the probability of $63\%$ for $\fpprob =0.01\%$ and almost always for higher error rates. These failure rates do not satisfy \reqlinky{lowfailure}.

\subsection{Membership at scale}
\label{sub:bio-practice}
We explore the potential of biometric fusion to lower the error rate, inspired by existing biometric identification systems such as Aadhar~\cite{Aadhaar} or US-VISIT~\cite{US-VISIT} that use multiple samples (10 fingerprint scans, 2 iris scans, and 1 facial image; and 10 fingerprints and 1 facial image, respectively) to lower their error rates when operating on a billion, respectively tens of millions, people.

The fusion procedure takes $f$ samples from different sources to create a $f$-ary template $\varAsList{X_i}_f$. As our primary goal is reducing FAR, our fusion requires all $f$ sample pairs of two templates to match as follows:
\begin{align*}
    M \in \{\binanswer\} \gets &\bioFusedMatch(\varAsList{X_i, Y_i, \dist_i, t_i, a_i}_f) =   \\
    &  \left\{\begin{array}{l}
        m_i \gets \bioAlignedMatch(X_i, Y_i, \dist_i, t_i, a_i) \\
        M \gets \textstyle \bigwedge_{i=1}^{f} m_i \text{.}\\
    \end{array}\right. 
\end{align*}
A fused false positive only happens when all $f$ underlying samples produce false positives at the same time. Therefore, the fused membership will have an FPR of $\prod_{i=1}^f {\fpprob}_i$ where ${\fpprob}_i$ is the FPR of the $i$'th sample. At the same time, the fused false negative rate is bound by $\sum_{i=1}^f {\fnprob}_i$.
Providing support for fusion is very similar to our alignment process ($\bioAlign$). The only difference is that when performing pairwise matching instead of performing the logical or ($\lor$) as in alignment, we use logical and ($\land$) in fusion. We support biometric fusion in all three instantiations of \name. This enables \name to scale its FAR rate to support very large userbases \emph{as long as \name receives the necessary number of biometric samples}.

\begin{figure}[t]
    \centering
    \includegraphics[width=\scalerFigSingleBio\linewidth, clip]{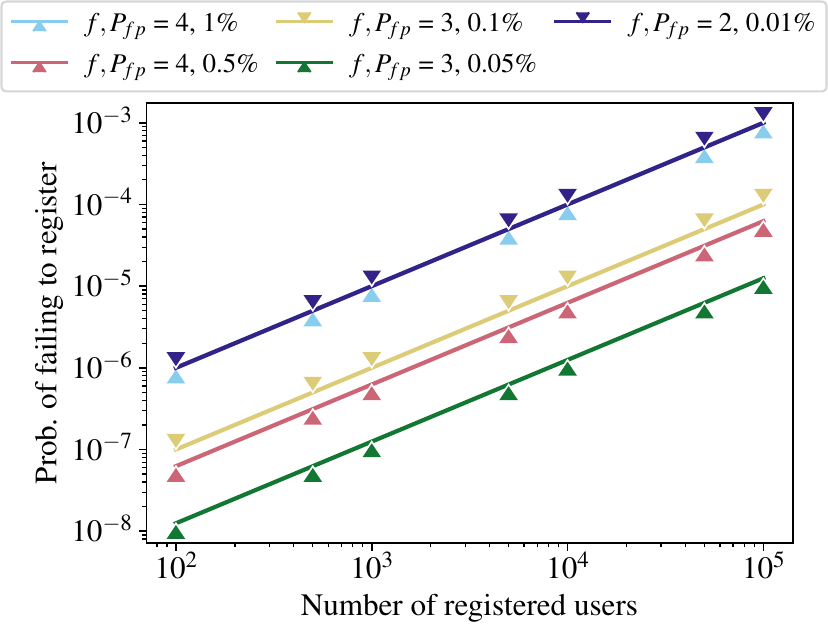}
    \caption{Impact of fusing $f$ samples with FPR of $\fpprob$.}
    \label{fig:far-multi-impact}
\end{figure}

We evaluate the FAR of membership when fusing $f$ samples with false positive rates of $\fpprob$ for various numbers of registered users in \cref{fig:far-multi-impact}. 
To satisfy \reqlinky{lowfailure}, we aim for a false accept rate of $\text{FAR}=\num{e-4}$ on deduplicating against a database of \num{10000} registered users. Two ways for achieving this error rate are: (1) taking 4 samples from biometric modalities with $\fpprob=1\%$ (e.g.,  4 fingers); or (2) with 2 samples with $\fpprob=0.01\%$ (e.g., 2 irises). Since there exist biometric devices that scan 4 fingers or 2 irises at the same time, \name could be efficiently deployed even when requiring more than one sample.
If lower error rates are desired, the ICRC can combine modalities (4 finger and 2 irises) or collect more finger samples from recipients.

\parasec{Template configuration} 
While we use existing biometric matchings as black boxes, they offer various configurations that provide accuracy/performance trade-offs that are of interest in \name. 
For example, having a larger template size (e.g., $\bioN=10k$ over $\bioN=2k$) or more alignments (e.g., $a=8$ over $a=1$) may improve both the false positive and negative error rates (up to a limit) at a performance cost.

Biometric fusion introduces the number of samples as a new trade-off dimension, where two short templates (e.g. $f\mathop{=}2, \bioN\mathop{=}2k$) might outperform and offer better error rates than a single large template (e.g. $f\mathop{=}1, \bioN\mathop{=}10k$).

\parasec{Failure rate of \name} \label{sub:bio-error}
The error rates of \name are mainly limited by the plaintext biometric matching. Our instantiations follow the plaintext computation with two minor differences:
\emph{Discretization:} SMC-\name and \shename expect biometric templates to have integer values from the domain $\Zx{d}$. While iris and finger codes follow this restriction, face templates are often real-valued and need discretization. 
\emph{Threshold:} The public threshold $t$ has limited precision.
As the encrypted computation is nearly identical to its plaintext counterpart, and measuring the accuracy of the underlying plaintext biometric approaches is out of the scope of this work, we do not measure the error rates of \name. \looseness=-1

\section{Evaluation}
\label{sec:eval}
\label{sub:implementation}
In this section, we evaluate the performance of \name. 

\begin{figure*}[t]
  \centering
  \includegraphics[width=1\linewidth, clip]{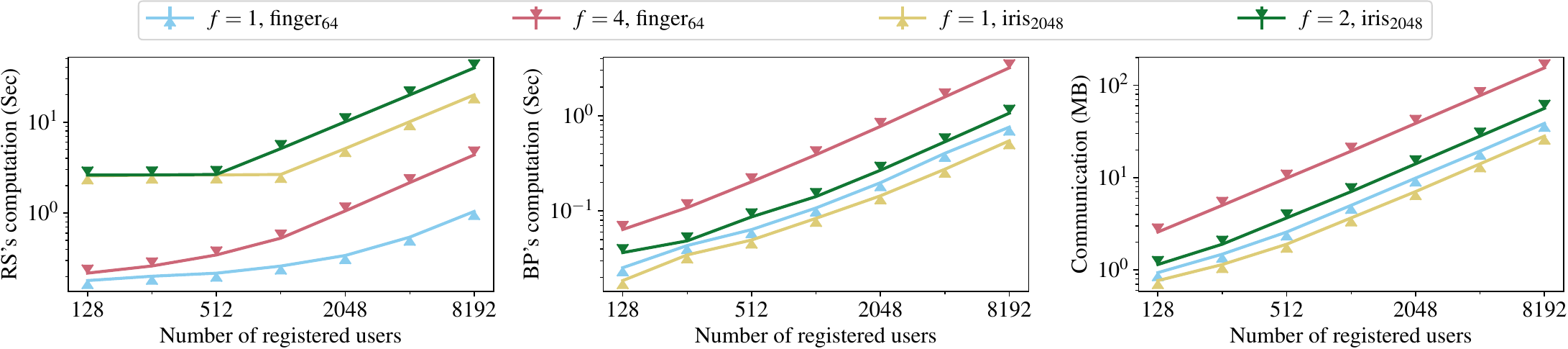}
  \caption{Evaluating the single-core membership performance of \shename with $f$ iris or fingerprint samples.}
  \label{fig:she-janus-eval}
\end{figure*}

\parasec{Biometrics setup} As we do not need to measure accuracy (see \cref{sub:bio-error}), we do not collect or process biometric data from real humans, and instead, take existing biometric template generation schemes and replace their output with random data of the same dimension as follows:

\para{Iris}  We base our iris code setting on templates produced by the University of Salzburg Iris Toolkit (USIT) v3~\cite{UIST-toolkit,rathgebUWH16}. 
The default parameters create $\bioN = 10240$ bits template, but smaller $2048$ bits templates are used in prior privacy-preserving works~\cite{BringerCP13}. Following our notation in \cref{sub:simplified-bio}, we create iris sensors as $\codify{Iris}_{10240} \gets \createBioSensor(10240, 2)$ and $\codify{Iris}_{2048} \gets \createBioSensor(2048, 2)$.

\para{Fingerprint} We use two instantiations of finger codes. The first sensor  $\codify{Finger}_{640} \gets \createBioSensor(640, 256)$ follows Jain et al.~\cite{JainPHP00} and uses Gabor filters to create a byte array of $\bioN = 640$ elements as the template. The second sensor $\codify{Finger}_{64} \gets \createBioSensor(64, 256)$ follows more recent CNN-based approaches\cite{EngelsmaJB22,EngelsmaCJ21} and creates $\bioN = 64$ byte templates. \looseness=-1

\para{Face} We base our face sensor  $\codify{Face} \gets \createBioSensor(512, 256)$ on ArcFace~\cite{DengGYXKZ22}. ArcFace generates 512-dimensional real-values templates, but we can discretize template elements into bytes to fit \name's requirements. Since face and fingerprint templates are similar and their matching follows the same process, we only evaluate fingerprints.

\parasec{Instantiations setup} We run our experiments on a machine with an Intel i7-8650U CPU and 16 GiB of RAM, using the following libraries and parameters for our instantiations.

\para{SMC-\name} We rely on a high-level SMC compiler called EMP-Toolkit~\cite{emp-toolkit} to generate and execute  garbled circuits. More specifically, we use `EMP-sh2pc'~\cite{emp-sh2pc} which offers semi-honest security in a two-party setting. We implemented SMC-\name fully in C++ (320 lines).

\para{\shename}
We use the BFV~\cite{FanV12} somewhat homomorphic encryption scheme implemented in the Lattigo library~\cite{lattigov4} for our homomorphic operations. Let the degree of the RLWE polynomial be $\polyDegree$, the plaintext modulus be $\plainMod$, and the ciphertext modulus be $\ctxMod$. We generate BFV keys with a set of parameters $(\polyDegree=4096, \lg(\ctxMod)=109, \plainMod=\num[group-separator = \,]{65929217})$. 
Our parameter set follows the homomorphic encryption security standard~\cite{Albrechtetal18} to provide 128 bits of security. Moreover, our parameter supports batching $\polyDegree$ scalar values into each ciphertext and performing SIMD operations. We implement \shename partially in Go (650 lines) and partially in C++ (150 lines).

\para{\teename} We use Intel SGX for our TEE-based solution and rely on the Fortanix enclave development platform~\cite{fortanix} to run Rust code in an SGX enclave. We implement \teename fully in Rust (300 lines). 

We will make all our implementations open source.

\subsection{Performance of \name}
We evaluate the performance of our three \name instantiations. We measure their (single-core) computation and communication cost when using various database sizes (generated as described above).
We assess both the single- and multi-sample settings with a single alignment. We use 4 fingers and 2 irises for the latter to ensure the instantiations satisfy \reqlinky{lowfailure} (see \cref{sub:bio-practice}). 

\para{\teename} 
Our TEE-based solution significantly outperforms \smcname and \shename, providing near plaintext biometric efficiency. Deduplication over an 8k user database takes under $\SI{50}{\milli\second}$ and uses less than $\SI{1}{\kibi\byte}$ of communication, meeting the scalability requirement (\reqlinky{scale}) easily.

\para{\shename} 
\cref{fig:she-janus-eval} displays the computation and communication costs for single and multi-sample membership across different database sizes using $\codify{Finger}_{64}$ and $\codify{Iris}_{2048}$ sensors. Deduplication 8k recipients with 2 irises takes $\SI{40}{\second}$ of computation on \rs and $\SI{56}{\mebi\byte}$, while deduplication with 4 fingerprints requires $\SI{4.4}{\second}$ and $\SI{155}{\mebi\byte}$. This duration, under a minute, is suitable for in-person registration, and \shename meets \reqlinky{scale}. When templates are larger, computation increases linearly, but not the communication cost (see \cref{app:evaluation-extra} for more details).\looseness=-1

\para{\smcname} 
\smcname offers information theoretic data security. With this instantiation, deduplicating \num{10000} users is prohibitively costly (see \cref{app:evaluation-extra}); it fails to meet \reqlinky{scale} and is better suited for smaller aid projects. Deduplicating 1k recipients with 2 irises requires $\SI{0.93}{\gibi\byte}$ and $\SI{39}{\second}$ of computation on both \rs and \bp, while deduplication with 4 fingerprints requires $\SI{52}{\second}$ and $\SI{2.36}{\gibi\byte}$.

\parasec{Comparison}
The \teename version is the most efficient but it requires SGX-supported devices at the \rsFull, as well as trust in the manufacturers of these devices. Also, the security of the scheme is limited by the existence of hardware vulnerabilities that might expose biometric data from the \rsFull.
While \smcname offers the strongest data protection, its cost makes it unsuitable for large aid distribution projects. \shename ensures biometric data safety using a quantum-secure encryption scheme, and registration completes in under a minute.
We would recommend \shename for large aid projects given its balance between security and performance; and \smcname as long as the size of the group receiving aid is such that performance is acceptable.

\subsection{Comparison with Closely Related Work}
\label{sub:comparison}
In this section, we compare the cost of \shename to two of the most relevant prior works:

\begin{figure}[t]
    \centering
    \includegraphics[width=0.8
    \linewidth, clip]{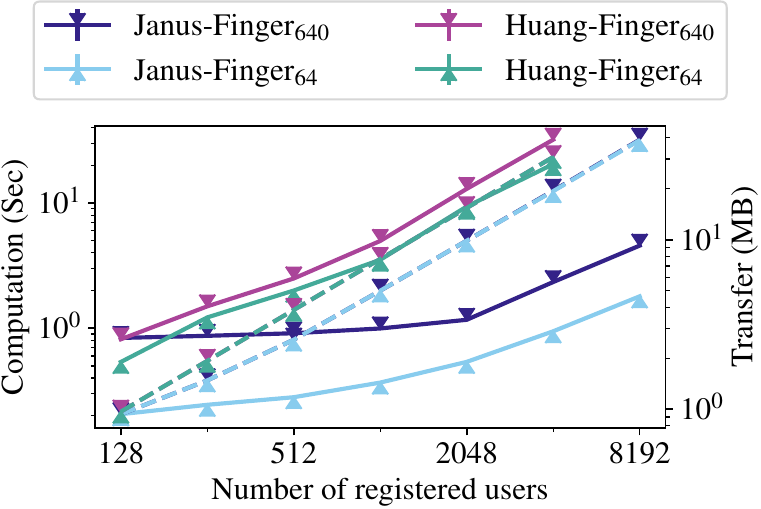}
    \caption{Single sample computation (solid lines) and communication (dashed lines) cost of Huang et al.~\cite{HuangMEK11} (80-bit security) vs \shename (128-bit security).}
    \label{fig:huang-eval}
\end{figure}

\parasec{Same protection} 
The closest paper to \name is Huang et al.~\cite{HuangMEK11}, which supports identification with a single finger code sample and has a structure similar to \shename. 
This work can be modified in a straightforward manner to support membership, but due to being limited to a single biometric sample and relying on the RSA assumption for security (which is not quantum secure), they do not satisfy \reqlinky{lowfailure} or \reqlinky{longterm}.  

Huang et al. use Pailler homomorphic encryption to compute the Euclidean distance of templates, then secret shares the distance between two parties, and finally uses a garbled circuit to determine the template with the highest similarity. Huang et al. have an expensive offline phase whose cost cannot be fully aggregated in our setting due to frequent user additions (see \reqlinky{dynamic}) and provides a protocol called `backtracking' to retrieve the identity of the matching template. We only consider the online cost of Huang et al. in the comparison without performing backtracking.
We run the code of Huang et al. on the machine described above using the original parameter of the paper that provides 80-bit (classic) security. \cref{fig:huang-eval} shows the cost of deduplicating over databases of various sizes with both $\codify{Finger}_{64}$ and $\codify{Finger}_{640}$ sensors. \smcname has a similar communication cost to Huang et al., but despite having a higher security guarantee of 128-bit, \name reduces the cost of deduplicating 4k users by a factor of 20x.

\begin{figure}[t]
    \centering
    \includegraphics[width=0.8\linewidth, clip]{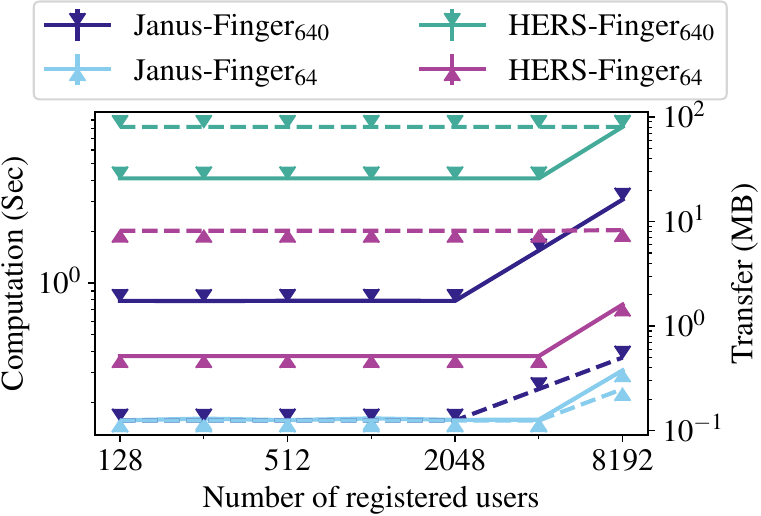}
    \caption{Single sample computation (solid lines) and communication (dashed lines) cost of computing template distance of HERS~\cite{EngelsmaJB22} and \shename (without thresholding).}
    \label{fig:hers-eval}
\end{figure}

\parasec{Less protection} HERS~\cite{EngelsmaJB22} provide privacy-preserving single sample fingerprint identification. HERS introduces a novel short finger code representation (the basis for our $\codify{Finger}_{64}$ sensor) and computes and \emph{reveals} the distance of templates using the BFV~\cite{FanV12} SHE scheme. HERS focuses on finding the most similar templates as part of identification and do not support or evaluate the accuracy of membership or deduplication. HERS is limited to a single biometric sample which prevents achieving a low membership error rate (\reqlinky{lowfailure}), reveals the distance against every template in the server's database that violates \reqlinky{single}, and does not offer a key rotation mechanism to protect against \reqlinky{pascomp} when actors are compromised at different times. \cref{fig:hers-eval} shows the cost of distance computation in \shename (without thresholding) and HERS. Both approaches have similar computation costs but \shename significantly improves the communication in our setting.

\section{Related work}
\label{sec:rw}
In \cref{sec:deduplication}, we discussed prior works related to the techniques we use to instantiate \name: secure multiparty computation, homomorphic encryption, and trusted execution environments; and discuss how they compare to our solution. We now discuss other techniques to build privacy-preserving biometrics systems.

\parasec{Fuzzy matching} 
Fuzzy matching approaches model the probabilistic biometric sampling process by assuming that a unique biometric ground truth exists, and taking each biometric sample read as a noisy version of this truth. 
Then, they decode each biometric reading using error correction codes. When biometrics samples are taken from the same user, the difference between them should be small, and decoding them should lead to the same base code.
There are three common fuzzy matching approaches: fuzzy commitments~\cite{JuelsW99, AdamovicMVSJ17, ImamverdiyevTK13}, that target biometric sources represented by fixed-size arrays; fuzzy vaults~\cite{Tams16, JuelsS06} which can support comparison of order-invariant arrays such as fingerprint minutiae; and fuzzy extractors~\cite{TongSLG07, DodisRS04}, that generate pseudorandom keys based on biometric samples.

The use of error correction codes negatively impacts accuracy, and thus fuzzy approaches cannot fulfill the low failure rate requirement (\reqlinky{lowfailure}). Also, many fuzzy matching protocols are vulnerable to statistical attacks~\cite{stoianov2009security, Stoianov10}, violating the single functionality requirement (\reqlinky{single}).

\parasec{Convolutional neural network} Besides using a CNN to learn a representation of biometric samples~\cite{EngelsmaCJ21}, it is possible to train a model to classify biometric images in a closed world~\cite{alwawiA2022}.  These approaches require retraining when adding a new recipient to the system, thus they do not satisfy the dynamic addition requirement \reqlinky{dynamic}.

\parasec{Biometric cryptosystems}
Biometric cryptosystems~\cite{RathgebU11} use biometric data to derive a secret key per user, either by deriving it from the biometric sample~\cite{TongSLG07} or by hiding the key in helper data which can be opened using a biometric sample from the user~\cite{JuelsS06, JuelsW99}. The secret key can be used to encrypt data or in authentication protocols. These approaches can be combined with so-called BioHashing~\cite{JinLG04, LuminiN07} that enhances the key derivation process by adding a user secret (password) to support 2-factor authentication and reduce the false accept rate. Deduplication cannot provide a secret key and thus biometric cryptosystems are not a suitable solution in our setting.

\section{Conclusion}
\label{sec:conclusion}
To ensure efficient distribution of aid, humanitarian organizations need to prevent recipients from registering more than once in their programs. 
Current approaches to achieving this are based on slow and not widely-available means. 
The humanitarian sector is turning to biometrics-based solutions to address these problems, but not carefully done these solutions can have a strong impact on the safety of aid recipients.

In this paper, we introduced \name, a biometric-based deduplication system designed to provide safety-by-design. \name can be combined with major biometric sources, and supports the fusion of multiple encrypted biometric samples to improve accuracy. We show that, while providing better protection, \name improves by orders of magnitude the performance of prior systems and, to the best of our knowledge, is the first to achieve the low error rate needed for deduplication in humanitarian aid. 

After Wang et al.'s system \cite{Pribad}, our work is a step forward toward enabling humanitarian organizations to harness digitalization to increase their efficiency while protecting the safety and dignity of those that are most in need. Further research is essential for safe digitalization in other stages, like pre-program recipient identification and post-aid monitoring and evaluation~\cite{mande}.

    \bibliographystyle{plain}
    \bibliography{sources}

    \appendices
    \crefalias{section}{appendix}
\section{Extended evaluation}
\label{app:evaluation-extra}
We evaluated the performance of \name in \cref{sec:eval}. In this section, we provide more details on the computation cost of \smcname and the impact of the template size on the performance.

\begin{figure*}[t]
    \centering
    \includegraphics[width=1\linewidth, clip]{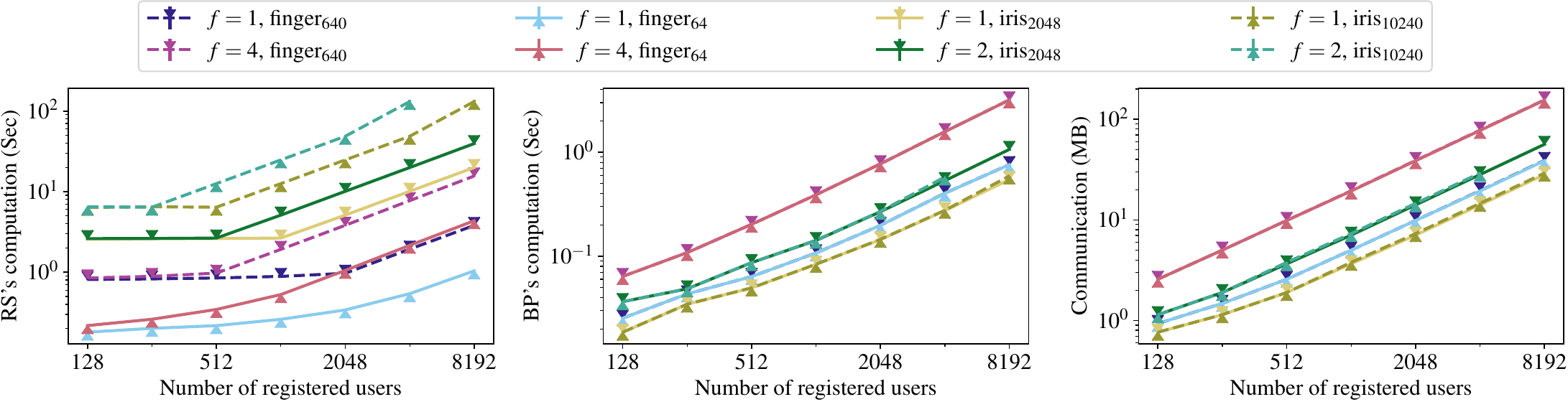}
    \caption{Evaluating the single-core membership performance of \shename with $f$ iris or fingerprint samples of varied sizes.}
    \label{fig:she-janus-large-eval}
\end{figure*}

\parasec{Template size} We measure the performance of deduplication with different template sizes in \cref{fig:she-janus-large-eval} to study the impact of template size on performance. We use solid lines to show shorter templates ($\codify{Finger}_{64}$ and $\codify{Iris}_{2048}$) and dashed lines to show longer templates ($\codify{Finger}_{640}$ and $\codify{Iris}_{10240}$). We observe that increasing the template size has a linear impact on the computation of the \rsFull, but does not impact the computation of \bpFull or the communication. This is in line with our expectation since the size of the template only impacts the HE computation of the distance while the communication and \bp's computation cost are dominated by the SMC thresholding.

\begin{figure*}[t]
    \centering
    \includegraphics[width=1\linewidth, clip]{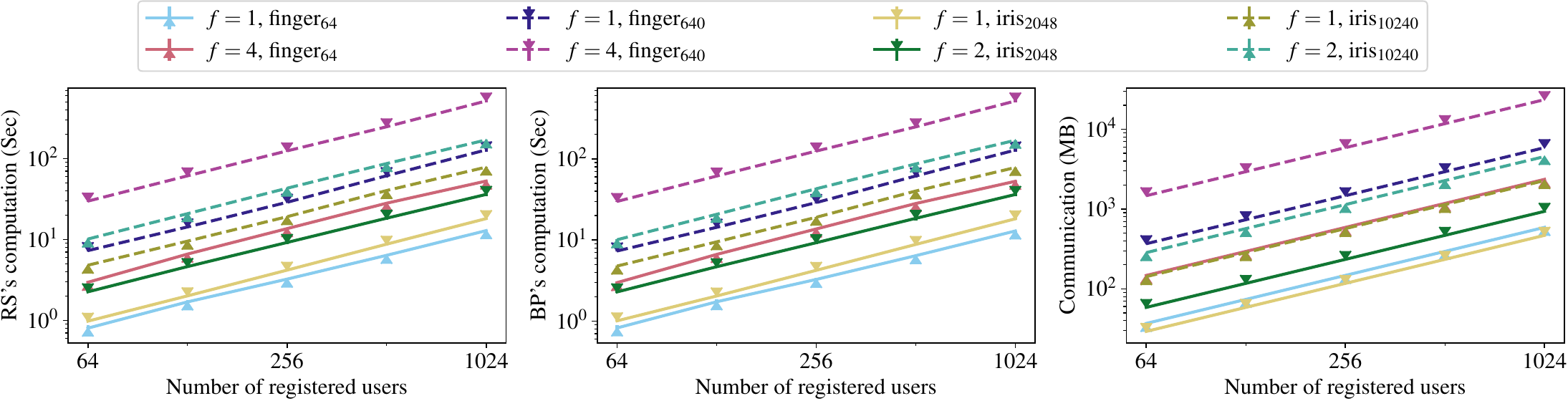}
    \caption{Evaluating the single-core membership performance of \smcname with $f$ iris or fingerprint samples over different database sizes.}
    \label{fig:smc-janus-eval}
\end{figure*}

\parasec{\smcname} \cref{fig:smc-janus-eval} shows the single-core computation and communication cost of operating \smcname with fingerprint and iris sensors of various sizes over different database sizes. Due to our use of garbled circuits, the computation cost of both parties (the \rsFull and \bpFull) are similar. Moreover, we observe that the number of biometric samples $f$, the template size $\bioN$, and the number of users registered in the database have a linear impact on all three \rs and \bp computation and communication costs.

\section{Normalized Hamming distance}
\label{app:normalized-hamming}
In this section, we define Normalized Hamming distance and discuss how we optimize our protocols to compute this distance.

\parasec{Noramlized Hamming distance}
A biometric template $X$ can have an associated binary mask $M_X$ to exclude certain values during comparison. The normalized Hamming distance, $\distNormHam((X, M_X), (Y, M_Y))$, applies both masks, calculates the Hamming distance, and then normalizes by the count of active comparison bits as follows:
$$\sum_i ((X[i] \lxor Y[i]) \land (M_X[i] \land M_Y[i])) * \bioN / |M_X \land M_Y| \text{.}$$

We apply the following adaptations to our instantiations:

\para{\teename}: Our TEE-based solution computes the matching in a plaintext domain inside an enclave where supporting normalize Hamming is trivial. 

\para{\smcname}: We store and handle the mask in the same manner as the biometric data (secret shared between two parties) in our SMC-based solution and adapt \cref{alg:smc-gc} to apply the masks and scale the final distance based on $\sum_i  (M_X[i] \land M_Y[i])$.

\para{\shename} While SHE schemes can support a multiplication depth beyond 1, increasing the depth impacts the schemes parameter selection leading to higher computation cost for each operation and a larger size for each ciphertext. We aim to have a multiplication depth of 1 in our \shename solution to be able to use the most efficient parameter with $\polyDegree=4096$. Using multiplication to compute both the masking $\land$ and the hamming $\lxor$ in the cipher domain leads to a minimum depth of 2. Therefore, we rely on a different representation of the hamming distance, namely  $\distHam(X, Y) = \sum_i (X[i] \land \lnot Y[i]) \lor  (\lnot X[i] \land Y[i])$, to compute the distance with a lower depth. When adding a new member, instead of separately encrypting and storing both $Y$ and $M_Y$, we store the following three vectors: $(\encvar{Y[i]\land M_Y[i]}, \encvar{\lnot Y[i] \land M_Y}, \encvar{M_Y})$. When performing the membership, we can compute $X[i]\land M_X[i]$ and $\lnot X[i]\land M_X[i]$ in the plaintext domain that allows us to compute the distance as $D \gets \sum_i (X[i]\land M_X[i])\cdot \encvar{\lnot Y[i]\land M_Y[i]} +  (\lnot X[i]\land M_X[i])\cdot \encvar{Y[i]\land M_Y[i]}$.

Note that while we use a SHE approach to compute the distance, our algorithms only require a linearly homomorphic scheme.

\end{document}